\documentclass{optica-article}
\usepackage{algorithmic}
\usepackage{algorithm}
\usepackage[title]{appendix}
\usepackage[section]{placeins}

\journal{opticajournal} % for journals or Optica Open

\articletype{Research Article}

\begin{document}

\title{Fully parallel optical matrix-matrix multiplication}

\author{Yufeng Zhang,\authormark{1} Hao Yan,\authormark{1} and Kaizhi Wang\authormark{1,*}}

\address{\authormark{1}School of Electronic Information and Electrical Engineering, Shanghai Jiao Tong University, Shanghai 200240, China}

\email{\authormark{*}kz\_wang@sjtu.edu.cn} %% email address is required; see note below about the corresponding author designation

% use {asbstract*} to suppress the copyright line. Copyright information will be added in production

\begin{abstract*} 
	In recent years, with the rapid development of electro-optic modulators, optical computing has become a potential excellent candidate for various computing tasks.
	New structures and devices for optical computing are emerging one after another, but the computing method is still the optical vector-matrix multiplication method that was decades ago.
	Here, we propose a novel optical computing paradigm that can parallelly implement matrix-matrix multiplication operation, which can directly replace existing vector-matrix multiplication, greatly improving computational efficiency.
    This preprint presents theoretical analysis, and we will supplement experimental results and conclusions in the future.
\end{abstract*}

%%%%%%%%%%%%%%%%%%%%%%%%%%  body  %%%%%%%%%%%%%%%%%%%%%%%%%%
\section{Introduction}
Matrix-matrix multiplication (MMM) is one of the core operations in computational and processing applications, widely used in signal processing, image processing, and deep learning.
For its $O(N^3)$ time complexity, MMM, which is composed of multiple vector-matrix multiplications (VMM), becomes the most time-consuming operation in various calculation task.
Due to the low performance of early computer technology, light has emerged as one of the ideal medium for replacing digital computing due to its excellent characteristics of low latency and low power consumption.
Since Dr. Goodman innovatively proposed the optical vector-matrix multiplications (OVMM) prototype \cite{ref1}, many researchers have utilized methods such as time multiplexing, wavelength multiplexing, and light source multiplexing to achieve more effective OVMM \cite{ref2,ref3,ref4,ref5,ref6,ref7,ref8,ref9,ref10}.
In addition to free space, Dr. Reck proposed a computational architecture based on Mach-Zehnder interferometer, verifying the possibility of calculations based on planar waveguides \cite{ref11}.
However, with the rapid development of integrated circuits and silicon based chip technology, optical computing, which is difficult to reconstruct structures and cannot update data in real-time, has no advantages compared to digital computing.
Therefore, the academic community has also paid more attention to the field of optical communication, resulting in the long-term stagnation of the development of optical computing.

In the 2010s, with the rise of technologies such as artificial intelligence and big data, massive MMM operations brought huge computational burden, and silicon based chips were unable to continue to meet computing needs.
At the same time, with the development of reconfigurable spatial light modulators (SLM) and photonic integrated circuits \cite{ref12,ref37,ref13}, optical computing is once again recognized as a potential high-performance computing solution \cite{ref14,ref15,ref16}.
Many researchers have implemented various forms of optical computing systems using free space and planar waveguides, and used them for training or inference of optical neural networks (ONNs) \cite{ref17,ref18,ref19,ref20,ref21,ref22,ref23,ref24,ref25,ref26,ref27,ref28,ref29}.
However, these systems still use decades ago OVMM methods, simply replacing traditional optical devices with higher speed, larger scale, and more reconfigurable advanced devices.
For example, using lenslet arrays or Dammann gratings instead of traditional multiple light sources to implement convolutional neural networks \cite{ref30,ref31,ref34}, using waveshapers and optical frequency combs instead of traditional four-wave mixing to achieve wavelength multiplexing computing \cite{ref20,ref22,ref32,ref35}, and using SLMs instead of traditional LED arrays and phase masks to achieve OVMMs \cite{ref17,ref27,ref33,ref36}.
Although the above works have greatly developed the equipment and combination strategies of optical computing, and effectively improved the actual performance of traditional optical computing methods, they have not fundamentally proposed more efficient optical computing principles.

Here, we innovatively proposed a fully parallelized optical matrix-matrix multiplication (POMMM) paradigm, which fundamentally changed Dr. Goodman's OVMM method.
The difference from previous methods is that our method does not require any pre-coding or pre-processing of matrixs, and the computing operation is completed through the propagation of light with single source and single wavelength.
This method has a simple and exquisite architecture, making it a universal computing method that is very suitable for accelerating ONNs and other optical computing operations.
POMMM adds an additional computational dimension to the ONN, enabling parallel training and inference of multiple samples and neurons.
In addition to OVMM and 2-D convolution operations, this method has the potential to become another new parallel computing paradigm for Fourier optics.

\section{Principle}
\subsection{Architecture of POMMM}
For MMM, assuming matrix $A$ (N rows, M columns) and matrix $B$ (M rows, N columns), then matrix $C=AB$ (N rows, N columns), the value $c_{nm}$ of the n-th row and m-th column of $C$ can be expressed as:
\begin{equation}
    c_{nm}=\sum_{i=1}^{M}{a_{ni}b_{im}}.
\end{equation}
Based on the above equation, we summarize the core steps of POMMM as (1) parallel implementation of row and column multiplication and addition (MAC) operations, and (2) moving the results to different positions, and design the architecture as shown in Fig.1.
\begin{figure*}[bp]
	\centering\includegraphics[width=13cm]{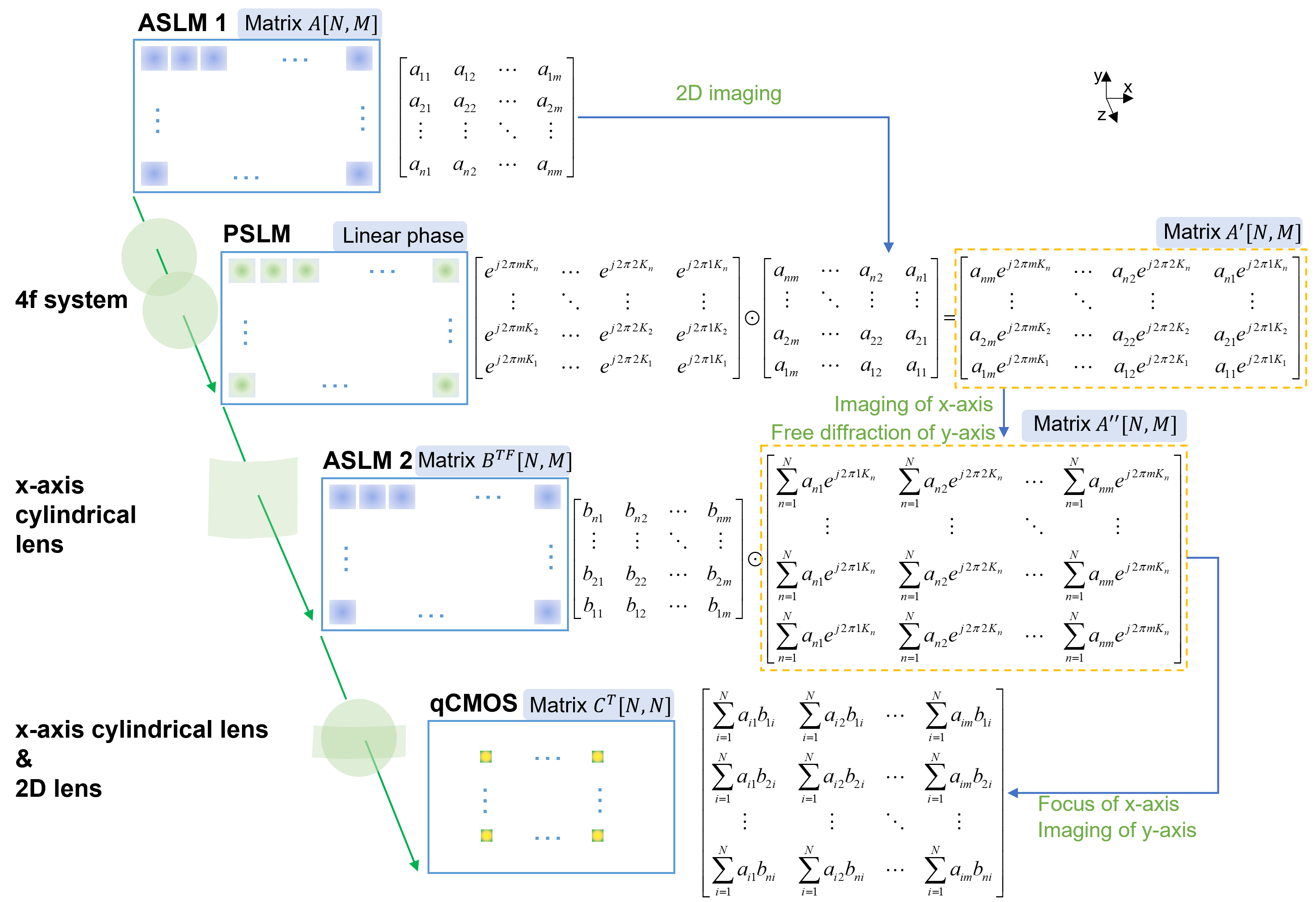}
	\caption{Basic principle. The optical architecture (lift panel) and the corresponding matrix operation (right panel). ASLM, amplitude spatial light modulator; PSLM, phase spatial light modulator; qCMOS, quantitative Complementary Metal Oxide Semiconductor camera.}
\end{figure*}
The matrix $A$ is modulated to the amplitude of wavefront by an amplitude spatial light modulator (ASLM), and is imaged to the surface of phase spatial light modulator (PSLM) by a 4f system.
PSLM will modulate a linearly changing phase along the x-direction (m) for each row of the matrix, and change its rate $K(n)$ along the y-direction (n) direction: $2\pi K(n)m$.
So the modulated wavefront can be represented by complex exponents, as shown in matrix $A'$ in Fig.1.
Then, the wavefront is imaged along the x-direction on the ASLM2 surface through a cylindrical lens, and the y-direction is the result of free diffraction.
Therefore, every point on the ASLM2 surface is a complex sum of matrix $A'$ along the y-direction, while the relationship in the x-direction remains unchanged, as shown in matrix $A''$ in Fig.1.
Transpose matrix $B$ and flip it along the y-direction to matrix $B^{TF}$, which is modulated to the wavefront through ASLM2 to achieve dot product ($\odot $) with the corresponding position of matrix $A''$:
\begin{equation}
    b^{TF}_{nm}\odot a''_{nm}=b^{TF}_{nm}\sum_{i=1}^{N}a_{im}e^{j2\pi K(i)m}.
\end{equation}
From the perspective of spatial frequency domain, the above equation indicates that each row of the matrix contains N spatial frequency components $K(1)~K(N)$, and the amplitude of the i-th frequency component on position (n,m) is the dot product of $a_{nm}$ and $b^{TF}_{nm}$.
By combining a 2-D lens with a x-direction cylindrical lens, imaging along y-direction and focusing (optical Fourier transform) along x-direction can be achieved, which is similar to traditional OVMM.
According to the principle of optical Fourier transform, when focusing along x-direction, N frequency components correspond to N focal points along x-direction, and the intensity of the n-th focal point is related to the sum amplitude of M points: $\sum_{i=1}^{M}a_{in}b^{TF}_{ni}$.
Fig.2 illustrates this process more vividly compared to traditional OVMM.
\begin{figure*}[bp]
	\centering\includegraphics[width=5cm]{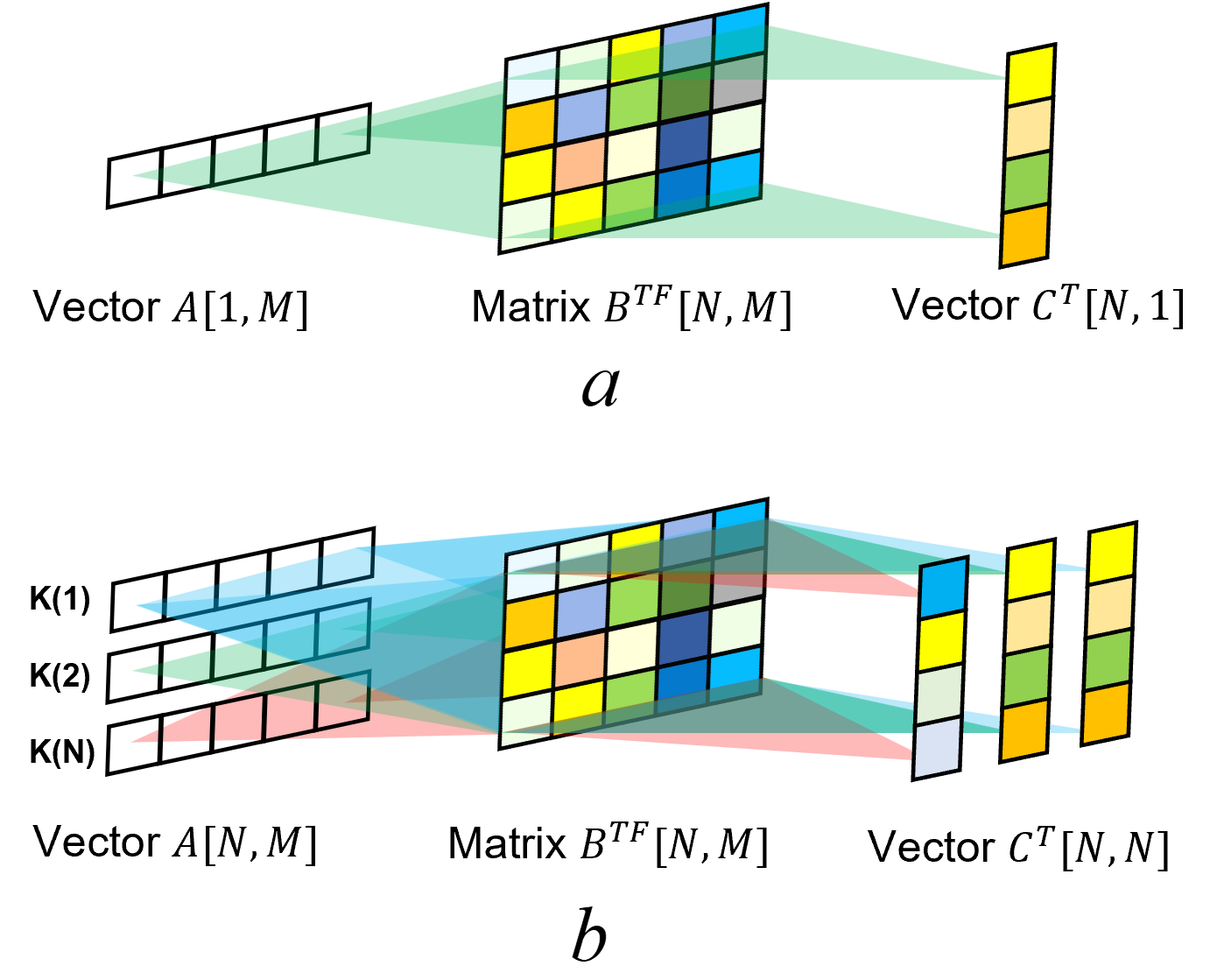}
	\caption{Comparison between POMMM and OVMM. (a) Principle of OVMM. (b) Principle of POMMM, different colors represent light modulated with different rates $K(n)$.}
\end{figure*}
Obviously, a $N \times N$-sized Matrix is composed by N frequency components in N rows captured by the qCMOS, which is the transposition of matrix $C$.

\subsection{ONN based on POMMM}
The most time-consuming part of neural network training and inference is the convolutional layer and fully connected layer, which are the parts that most ONNs attempts to accelerate.
The principles of convolutional layer and fully connected layer are both based on VMM.
As shown in Fig.3, the convolutional layer is to achieve the inner product of the slices with the convolutional kernels to form a new sample, the fully connected layer is to achieve VMM on samples and weight matrix ($\omega_nm$ and $\omega_ij$).
For convolutional layers, POMMM can process multiple convolutional kernels ($k_m1$,$k_m2$,...) in parallel, which is very effective for multi feature extraction.
The fully connected layer based on POMMM can process multiple samples ($a_1m$,$a_2m$,...) in parallel, greatly improving computational speed.
By comparison, simply replacing the existing OVMM with our POMMM can achieve parallel convolutional layer and fully connected layer processing, greatly improving training and inference of ONNs.
\begin{figure*}[bp]
	\centering\includegraphics[width=13cm]{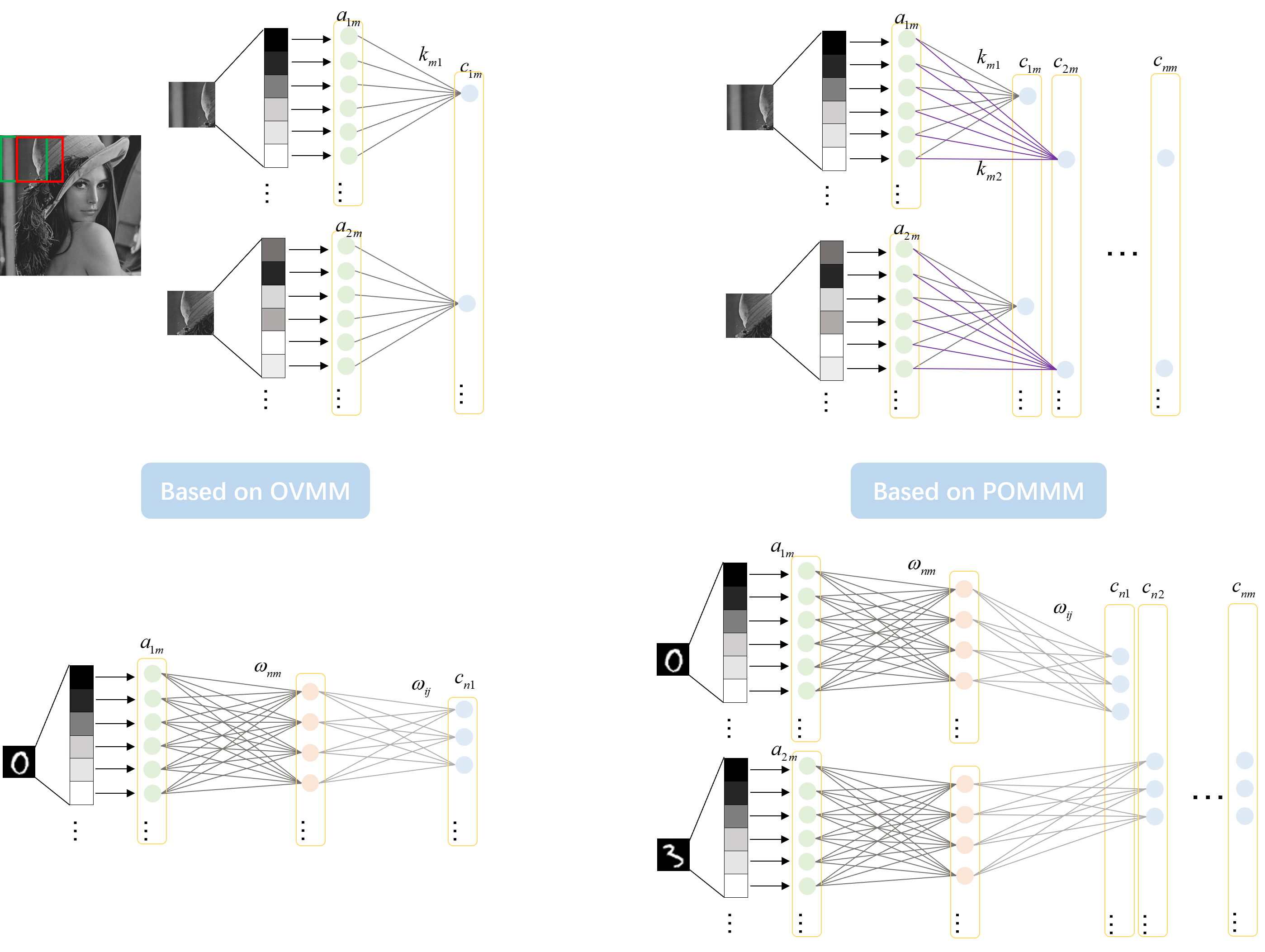}
	\caption{ONN based on POMMM. The convolutional layer (up panel) and the fully connected layer (down panel). The life part is based on traditional OVMM, while the right part is based on our POMMM.}
\end{figure*}

%%%%%%%%%% If using BibTeX:
\bibliography{reference}

%%%%%%%%%% If preparing manually:
% \begin{thebibliography}{1}
% \newcommand{\enquote}[1]{``#1''}

% \bibitem{Zhang:14}
% Y.~Zhang, S.~Qiao, L.~Sun, Q.~W. Shi, W.~Huang, L.~Li, and Z.~Yang,
%   \enquote{Photoinduced active terahertz metamaterials with nanostructured
%   vanadium dioxide film deposited by sol-gel method,}
%   {\protect\JournalTitle{Optics Express}} \textbf{22}, 11070--11078 (2014).

% \bibitem{Optica}
% {Optica}, \enquote{{Optica Publishing Group},}
%   \url{http://www.opg.optica.org}.

% \bibitem{FORSTER2007}
% P.~Forster, V.~Ramaswamy, P.~Artaxo, T.~Bernsten, R.~Betts, D.~Fahey,
%   J.~Haywood, J.~Lean, D.~Lowe, G.~Myhre, J.~Nganga, R.~Prinn, G.~Raga,
%   M.~Schulz, and R.~V. Dorland, \enquote{Changes in atmospheric consituents and
%   in radiative forcing,} in \enquote{Climate Change 2007: The Physical Science
%   Basis. Contribution of Working Group 1 to the Fourth assesment report of
%   Intergovernmental Panel on Climate Change,}  S.~Solomon, D.~Qin, M.~Manning,
%   Z.~Chen, M.~Marquis, K.~B. Averyt, M.~Tignor, and H.~L. Miler, eds.
%   (Cambridge University Press, 2007).

% \end{thebibliography}

\end{document}